\documentclass[preprint,12pt]{elsarticle}

\begin{document}

\begin{frontmatter}

\title{Observational analysis of bulk viscous modified Chaplygin gas in (2+1)-dimensional universe using MCMC}

\author[label1]{Praveen Kumar Dhankar\footnote{pkumar6743@gmail.com}} 
\affiliation[label1]{organization={Symbiosis International (Deemed University)},
            addressline={Nagpur Campus}, 
            city={Nagpur},
            postcode={440008}, 
            state={Maharastra},
            country={India}}

\author[label2]{Aritra Sanyal\footnote{aritra@iases.org.in}} 
\affiliation[label2]{organization={Institute of Astronomy Space and Earth Science (IASES)},
            addressline={P 177, CIT Road, Scheme 7m}, 
            city={Kolkata},
            postcode={700054}, 
            state={West Bangal},
            country={India}}

\author[label3]{Albert Munyeshyaka\footnote{munalph@gmail.com}} 
\affiliation[label3]{organization={Rwanda Astrophysics, Space and Climate Science Research Group, University of Rwanda},
            addressline={College of Science and Technology}, 
            city={Kigali},
            country={Rwanda}}

\author[label4]{Saibal Ray\footnote{saibal.ray@gla.ac.in}} 
\affiliation[label4]{organization={Center for Cosmology, Astrophysics and Space Science (CCASS), GLA University},
            city={Mathura},
            postcode={281406}, 
            state={ Uttar Pradesh},
            country={India}}

\author[label5,label6,label7,label8]{Behnam Pourhassan\footnote{b.pourhassan@du.ac.ir}} 
\affiliation[label5]{organization={Center for Theoretical Physics, Damghan University},
            city={Damghan},
            postcode={3671641167}, 
            country={Iran}}
\affiliation[label6]{organization={School of Physics, Khazar University},
              city={Baku},
              country={Azerbaijan}}
\affiliation[label7]{organization={Centre for Research Impact \& Outcome},
            addressline={Chitkara University Institute of Engineering and Technology, Chitkara University}, 
            city={Rajpura},
            postcode={140401}, 
            state={Punjab},
            country={India}}
\affiliation[label8]{organization={Canadian Quantum Research Center},
            city={204-3002 32 Ave Vernon},
            postcode={BC V1T 2L7}, 
            country={Canada}}



\begin{abstract}
This paper investigates regarding cosmological implications of a bulk viscous modified Chaplygin gas (MCG) in (2+1)-dimensional Friedmann-Robertson-Walker spacetime, incorporating both theoretical analysis and observational constraints. We derive analytical solutions for both viscous and non-viscous cases, revealing distinct behavior in energy density evolution, Hubble parameter dynamics, and deceleration parameter transitions. A comprehensive perturbation analysis illustrates how bulk viscosity dampens the structure growth oscillations, addressing a key challenge faced by Chaplygin gas models in higher dimensions. Using  Markov chain Monte Carlo (MCMC) techniques with Hubble parameter and Pantheon supernova datasets, we impose constraints on our model parameters, obtaining $H_0 = 67.90$ km s$^{-1}$ Mpc$^{-1}$, showing remarkable consistency with Planck $\Lambda$CDM estimations despite the dimensional reduction. Our findings suggest that lower-dimensional viscous cosmology captures essential features of cosmic evolution while providing valuable theoretical insights into the interplay between dissipative effects and exotic equations of state. 
\end{abstract}


\end{frontmatter}


\section{Introduction}
General Theory of Relativity (GTR or simply GR) in $(2+1)$-dimensional spacetime is regardred to possess a number of special  features which simplifies the studying problem. As revealed from the works of severel scientists that no gravitational waves as well as no black holes in the absence of negative cosmological constant, the Weyl curvature is identically zero and weak field limit of the theory does not correspond to Newtonian gravity in two space dimensions \cite{Giddings,Barrow,Staruszkiewicz,Gott,Deser1,Deser2,Deser3,Deser4,BTZ}. Of course here `simplifying features' means that considerable progress can be made in search of the general cosmological solution to the three dimensional Einstein equations. It has been observed that the cosmological solution are rather cumbersome and dominated by non-integrability, whereas the theory in $(2+1)$-dimension offers the possibility of finding the general solutions \cite{Staruszkiewicz}. The status and attributes of the Einstein field equations in two spatial and one temporal dimensions provide the unique reason to focus on $(2+1)$-dimension \cite{1,2,3}. Because of the reason that the Einstein and Riemann tensors are equivalent in  $(2+1)$-dimension, so the spacetime is flat outside the sources where there no gravitational field does exist and no Newtonian limit also is applicable.

The unique relationship between the Riemann and Ricci tensors in 3-dimensional spacetime reveals a fascinating peculiarity of Einstein's general relativity in lower dimensions. Unlike in 4-dimensional spacetime, where the Riemann curvature tensor carries additional information beyond the Ricci tensor through the Weyl tensor, in 3-dimensional case the Riemann tensor can be completely determined via the Ricci tensor. This mathematical constraint has profound physical implications: any solution to the vacuum Einstein equations (where the Ricci tensor vanishes) must necessarily describe flat spacetime, except at possible singular points. This striking feature therefore means that gravitational waves cannot propagate through empty space in 3-dimensions, as there are no {\it free} gravitational degrees of freedom. This unique characteristic makes a 3-dimensional gravity an invaluable theoretical laboratory for exploring fundamental concepts in gravity, despite its obvious departure from our 4-dimensional reality. The absence of propagating gravitational waves, however, does not mean that 3-dimensional gravity is trivial. It has been shown in ref. \cite{Clement} that particularly interesting physics emerges when one considers stationary solutions, especially in the presence of matter content or an erstwhile cosmological constant. These solutions can exhibit rich geometric structures, including black holes and cosmic strings, though their properties often differ markedly from their 4-dimensional counterparts. The study of such solutions has proven invaluable for understanding quantum gravity, holography, and the AdS/CFT correspondence \cite{maldacena}. This special property of 3-dimensional gravity also highlights the dimensional dependence of gravitational physics, demonstrating how fundamental aspects of gravity can change dramatically with the number of spacetime dimensions. This insight has motivated the study of gravity in various dimensions.

It has been investigated that all hydrostatic structures in $(2+1)$-dimensional GR contain matter-filled spaces without any matching to the external vacuum solution and thus represent a specific static cosmology \cite{Cornish}. Interestingly, several cosmological observations specify that there must be some kind of dark energy with a repulsive pressure in the late-time universe. Therefore, it has created interest to the study of cosmological scaling solutions of minimally coupled scalar fields in 3-dimension \cite{Martinez}. Fujiwara et al. \cite{Fujiwara}  have explained that the case of nucleation of the universe in a $(2 + 1)$-dimensional gravity model with a negative cosmological constant. 

In connection to the above discussion, an interesting model to describe the dark energy is Chaplygin gas (CG) \cite{Kamenshchik,Betnto} which emerged initially in cosmology from the string theoretical concept  \cite{Barrow1,Barrow2} which are based on CG equation of state (EOS) and developed to the generalized Chaplygin gas (GCG) \cite{Bilic}. Further, GCG was extended to modified Chaplygin gas (MCG) \cite{Debnath}. The CG and its various modifications have emerged as significant models in modern cosmology, offering potential explanations for the universe's accelerated expansion and dark energy behavior. Early investigations focused on the Friedmann-Lemaître-Robertson-Walker (FLRW) bulk viscous cosmology incorporating MCG in flat space \cite{4}, while subsequent studies explored viscous effects in GCG models \cite{3,4}. The framework was further extended to include space curvature and cosmological constant considerations \cite{7,8}, with particular attention given to the interplay between shear and bulk viscosities \cite{9,10}. The model's versatility led to its application in string theory through interactions with closed string tachyons \cite{11} and its role in inflationary scenarios \cite{12,13,14}. More sophisticated developments incorporated varying gravitational and cosmological constants \cite{15}, holographic implications \cite{16}, phantom cosmology \cite{17}, and inflationary models \cite {18}. An interesting extension of CG model which include all mentioned effects is called extended Chaplygin gas (ECG) \cite{19}. 

Recent advances have expanded the model's applications to Horava-Lifshitz gravity \cite{20}, sound speed analysis \cite{21}, and black hole physics in Anti-de Sitter space \cite{22}, demonstrating the ECG model's continued relevance in contemporary theoretical physics. The aspect of modified Chaplygin gas (MCG) have studied by many researchers with $(2+1)$-dimensional spacetime \cite{Khadekar,Kumar} as well as in the anisotropic universes to perform a comparative study of the Kantowski–Sachs and Bianchi-I models \cite{Ray} and on the brane \cite{Paul}. While Chaplygin gas models have garnered significant attention as candidates for unified dark matter-cum-dark energy scenarios, it is important to acknowledge the observational challenges these models have faced in recent years. Several studies have highlighted tensions between predictions of Chaplygin gas models and observational constraints in the standard $(3+1)$-dimensional framework.

The structure formation data have shown that the original CG and some of its modifications face difficulties in simultaneously satisfying constraints from cosmic microwave background radiation (CMBR), baryon acoustic oscillations (BAO), and large-scale structure observations \cite{1007.1011, 1212.4136}. These tensions arise primarily from the behavior of perturbations in Chaplygin gas models, which tend to produce oscillations or instabilities in the matter power spectrum that are not observed in galaxy surveys. Furthermore, recent results from the Dark Energy Spectroscopic Instrument (DESI) have provided new constraints on dark energy models. While these observations have revitalized discussions about the nature of dark energy, they have not specifically favored Chaplygin gas formulations over other dark energy models \cite{2308.15776}. Some analyses suggest that unless an additional scalar field carrying vacuum energy is introduced into Chaplygin gas models, they remain disfavored particularly at early times. Despite these challenges in the standard cosmological framework, studying Chaplygin gas models in alternative dimensional settings remains valuable for several reasons. First, the mathematical properties of these models in simplified frameworks like $(2+1)$-dimensions can provide insights into their fundamental behavior. Second, exploring the effects of bulk viscosity in these models addresses theoretical questions about dissipative effects in cosmic fluids that transcend specific model implementations. Finally, the analytical tractability of lower-dimensional models allows for more precise examination of specific features like the transition between different cosmic epochs.

Our present work should therefore be understood as a theoretical investigation into the properties of bulk viscous modified Chaplygin gas in a simplified dimensional framework, rather than a proposed alternative to the standard cosmological model. The insights gained from this analysis may inform future developments in cosmic fluid models and contribute to our understanding of dissipative effects in cosmological settings. In particular, the $(2+1)$-dimensional framework offers unique advantages for theoretical cosmology. The Einstein field equations in $(2+1)$-dimensional spacetimes exhibit a direct relationship between the Riemann and Ricci tensors, eliminating the complications arising from the Weyl tensor present in higher dimensions. This mathematical simplification allows for more tractable analysis of cosmic fluids' behavior without the additional complexity of gravitational waves or certain non-linear effects present in $(3+1)$-dimensions. The $(2+1)$-dimensional approach has proven valuable in quantum gravity research as a theoretical laboratory where fundamental concepts can be explored with greater clarity \cite{Carlip2003}. In this context, our study of bulk viscous MCG in $(2+1)$-dimensions provides a controlled environment to isolate and analyze the interplay between viscosity and the exotic equation of state, potentially revealing dynamics that might be obscured in more complex dimensional settings. Furthermore, the analytical solutions obtained in lower dimensions can guide numerical approaches in physically realistic (3+1)-dimensional scenarios, particularly regarding the transition phases between a decelerated to accelerated expansion of the universe. 

The study of cosmological models in $(2+1)$-dimensions, while not directly corresponding to our observable universe, has profound theoretical significance in contemporary physics. The $(2+1)$-dimensional spacetime serves as an important theoretical laboratory for testing fundamental aspects of gravitational theories without the full complexity of $(3+1)$-dimensions \cite{Carlip2003, Witten2007}. In these lower-dimensional models, many of the technical difficulties associated with quantum gravity and non-linear effects are significantly reduced, allowing for exact solutions that are often unattainable in the physically realistic $(3+1)$-dimensional case. This dimensional reduction approach has precedent across physics, from the study of lower-dimensional quantum field theories to condensed matter systems, where simplification often reveals essential underlying mechanisms that remain operative in higher dimensions. Particularly for viscous cosmic fluids like those studied here, the $(2+1)$-dimensional framework allows us to isolate the effects of bulk viscosity on cosmic evolution without the additional complexities introduced by shear viscosity and gravitational wave perturbations present in higher dimensions. The insights gained from these simplified models--especially regarding phase transitions and acceleration mechanisms--can inform and guide more complex numerical investigations in $(3+1)$-dimensional cosmology. Thus, while our $(2+1)$-dimensional analysis is not proposed as a direct model of the observable universe, it provides valuable theoretical insights into the fundamental behavior of cosmological fluids that may translate to physically realistic scenarios.

Our motivation in the ongoing work is mainly to focus on constraining the model parameters by using various observational datasets as available recently. In particular, our investigating scheme is as follows: (i) we have chosen to use $H(z)$ dataset consisting of 30 measurements and Pantheon dataset consisting of 1048 measurements to get the best fit values of the model parameters, (ii) we adopt Markov chain Monte Carlo (MCMC) analysis to fix the model parameters on the recently released data, (iii) we execute observational data analysis by $\chi^2$-minimization technique in view of $H(z)$+SNIa dataset \cite{Gadbail} which provides us the bounds of arbitrary parameters within $1\sigma, 2\sigma$ confidence levels. 

The outline of the present investigation is envisaged as follows: In Section 2, we have obtained Einstein filed equations for flat FRW cosmological models and MCG in the framework of $(2+1)$-dimensional spacetime. In Section 3, the model parameters are constrained by using the Hubble and Pantheon datasets. Concluding remarks are given in Section 4.

\section{FRW Model and Friedmann equations}

We consider the FLRW universe in $(2+1)$-dimensional spacetime as \cite{Khadekar}
\begin{eqnarray}\label{1}
	ds^{2}= -dt^{2}+a^{2}(t)\bigg[\frac{dr^{2}}{1-kr^{2}}+r^{2}d\theta^{2}\bigg],
\end{eqnarray}
where $a(t)$ represents the scale factor. The coordinates ($t$,~$r$,~$\theta$) represent the co-moving coordinates and the constant $k$ denotes the curvature of the space $k=0,1,-1$ for flat, closed and open universe respectively.

The Einstein field equations in $(2+1)$-dimension spacetime can be written as
\begin{eqnarray}\label{2}
	G_{ij}=R_{ij}-{\frac{1}{2}}Rg_{ij}= 2{\pi}GT_{ij},
\end{eqnarray}
where $G_{ij}$ is the Einstein tensor, $R_{ij}$ is the Ricci tensor, $R$ is the Ricci scalar. 

The energy momentum tensor corresponding to the bulk viscous fluid is given by \cite{Saadat}
\begin{eqnarray}\label{3}
	T_{ij}=(\rho+\bar{p})u_{i}u_{j}-\bar{p}g_{ij}, 
\end{eqnarray}
where $\rho$ is the energy density and $u^{i}$ is the velocity three vector with $u^{i}u_{j} = -1$. 

The total pressure and the proper pressure involve bulk viscosity coefficient $\xi$ and Hubble expansion parameter $H = \dot{a}/a$ are given by the following equations:
\begin{eqnarray}\label{4}
	\bar{p}=p-2{\xi}H, 
\end{eqnarray}
and modified Chaplygin gas EOS can be written as
\begin{eqnarray}\label{5}
	p={\gamma}{\rho}-\frac{A}{\rho^{\beta}}, 
\end{eqnarray}
with $A>0$ and $0<\beta \leq 1$ where $\gamma$ and $A$ describes the features of dark energy models and Chaplygin gas respectively. 

The modified Chaplygin gas equation of state (\ref{5}) represents a cosmic fluid with remarkable transitional properties. The parameter $\gamma$ controls the fluid's behavior at high densities (early universe), where the term $\gamma\rho$ dominates. In this regime, the MCG approximately behaves as a perfect fluid with equation of state $p = \gamma\rho$. When $\gamma > 0$, this corresponds to conventional matter or radiation, while $\gamma < 0$ would represent phantom-like behavior. As the universe expands and density decreases, the second term $-A/\rho^\beta$ gradually becomes more significant. This term introduces negative pressure, driving the accelerated expansion characteristic of dark energy. The transition between these two regimes occurs naturally as the universe evolves, allowing a single fluid component to mimic the behavior of both dark matter and dark energy—a conceptual simplification compared to the standard cosmological model's separate components. The introduction of bulk viscosity (\ref{4}) represents energy dissipation within the cosmic fluid. While perfect fluids in standard cosmology expand and contract adiabatically, real physical systems generally exhibit dissipative effects. Bulk viscosity accounts for the entropy production associated with the homogeneous expansion of the universe, modeling departures from equilibrium during cosmic evolution. Physically, the bulk viscosity coefficient $\xi$ quantifies the fluid's resistance to uniform expansion or contraction. When the universe expands, positive bulk viscosity generates an effective negative pressure (beyond what the equation of state provides), potentially enhancing the acceleration effect. This dissipative mechanism has no analog in standard perfect fluid cosmology, offering an alternative physical process that could contribute to cosmic acceleration without invoking exotic forms of energy.

The field equations (\ref{2}) with the help of line element (\ref{1}) for flat space universe in $(2+1)$ dimensions are given by \cite{Betnto}
\begin{eqnarray}\label{6}
	{\frac{\dot{a}^2}{{a}^2}}=\frac{\rho}{2},
\end{eqnarray}
and
\begin{eqnarray}\label{7}
	\frac{\ddot{a}}{a}=-\bar{p},
\end{eqnarray}
where a $(^.)$ means the derivative with respect to the cosmic time $t$. 

The modified Einstein field equations in $(2+1)$-dimensions  reflect fundamental differences from their $(3+1)$-dimensional counterparts. In $(2+1)$-dimensions, spacetime curvature is entirely determined by the distribution of matter, with no propagating gravitational degrees of freedom. This means that the gravitational field exists only where matter is present--a stark contrast to $(3+1)$-dimensions where gravitational waves can propagate through empty space. Physically, Eq. (\ref{6}) relates the expansion rate (represented by the Hubble parameter) directly to the energy density, while Eq. (\ref{7}) connects the acceleration of the scale factor to the effective pressure. The absence of spatial curvature terms (when $k=0$) simplifies the dynamics, allowing us to isolate the effects of the exotic fluid without the additional complexity of curved space geometry.

The energy-momentum conservation equation in $(2+1)$-dimensional spacetime can be supplied as
\begin{eqnarray}\label{8}
	\dot{\rho}+2\frac{\dot{a}}{a}(\rho+\bar{p})=0.
\end{eqnarray}

By using Eqs. (\ref{4})--(\ref{6}) in the conservation Eq. (\ref{8}), we get the reduced form as follows:
\begin{eqnarray}\label{9}  
{\dot{\rho}}+\sqrt{2}{(\gamma+1)}\rho^{3/2}-2{\xi}{\rho}-\sqrt{2}A=0.
\end{eqnarray}

We solve this equation for two different cases: namely, $\xi = 0$ and $\xi \neq 0$ (constant).\\

{\bf{Case (i): when $\xi = 0$}}\\

In this case from Eq. (\ref{9}), we simply obtain    
\begin{eqnarray}\label{10} 
	\dot{\rho}+\sqrt{2}{(\gamma+1)}\rho^{3/2}=\sqrt{2}A.
\end{eqnarray}

After integrating this equation, we get
\begin{eqnarray}\label{11} 
	{\rho(a)}=\bigg[\frac{1}{(\gamma+1)}\big(A-\frac{c}{a^{3(\gamma+1)}}\big)\bigg]^{2/3},
\end{eqnarray}
where $c$ is a constant of integration.

By using the value of $\rho$ from Eq. (\ref{11}) in Eq. (\ref{6}), we get the Hubble parameter $H$ as
\begin{eqnarray}\label{12} 
	{H(a)}={\frac{1}{\sqrt{2}}}\bigg[\frac{1}{(\gamma+1)} \big(A-\frac{c}{a^{3(\gamma+1)}}\big)\bigg]^{1/3}.
\end{eqnarray}

Applying the transformation with $a = (1 + z)^{-1}$, on the above expression for energy density $\rho$ and Hubble parameter $H$ in terms of redshift $z$ can be re-write of Eqs. (\ref{11}) and (\ref{12}) as
\begin{eqnarray}\label{13} 
	{\rho(z)}=\bigg[\frac{1}{(\gamma+1)}\big(A- c.(1 +z)^{3(\gamma+1)}\big)\bigg]^{2/3},
\end{eqnarray}

\begin{eqnarray}\label{14} 
	{H(z)}={\frac{1}{\sqrt{2}}}\bigg[\frac{1}{(\gamma+1)} \big(A- c. {(1+z)^{3(\gamma+1)}}\big)\bigg]^{1/3},
\end{eqnarray}

\begin{figure}[htbp]
	\centering
	\includegraphics[scale=0.40]{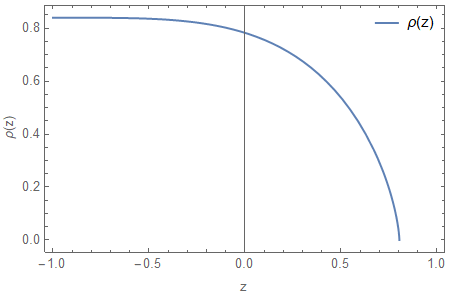}
	\caption{The energy density $\rho(z)$ is shown against redshift $z$ for different values of the constants $\gamma=0.3, A=3.4, c=1$ for the Case (i).}
\end{figure}

\begin{figure}[htbp]
	\centering
	\includegraphics[scale=0.60]{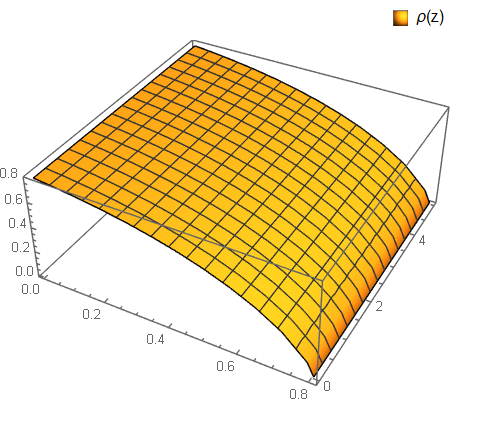}
	\caption{The energy density $\rho(z)$ is shown against redshift $z$ in the contour plot for the different values of $\gamma=0.3, A=3.4, c=1$ for the Case (i).}
\end{figure}

\begin{figure}[htbp]
	\centering
	\includegraphics[scale=0.40]{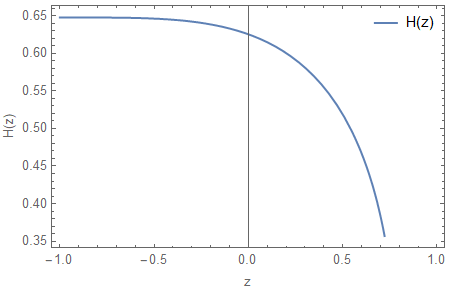}
	\caption{The Hubble parameter $H(z)$ is shown against redshift $z$ for different values of the constants $\gamma=0.3, A=3.4, c=1$ for the Case (i).}
\end{figure}

\begin{figure}[htbp]
	\centering
	\includegraphics[scale=0.60]{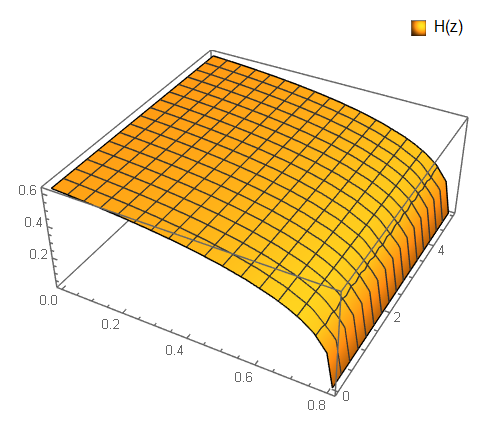}
	\caption{The Hubble parameter $H(z)$ vs redshift $z$ in the contour plot for the different values of $\gamma=0.3, A=3.4, c=1$ for the Case (i).}
\end{figure}

\begin{figure}[htbp]
	\centering
	\includegraphics[scale=0.40]{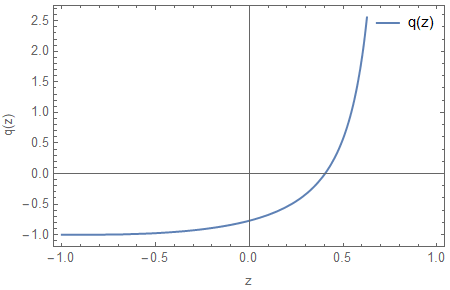}
	\caption{The deceleration parameter $q(z)$ is shown against redshift $z$ for different values of the constants $\gamma=0.3, A=3.4, c=1$ for the Case (i).}
\end{figure}

From Eq. (\ref{12}), it is observed that the energy density $\rho$ is decreases when the red shift $z$ is increases which is shown in Figs. 1 and 2. From Eq. (\ref{14}), we observe that the Hubble parameter  $H(z)$ is increases when the scale factor $z$ is increases  and approaches to an infinitesimal constant at late time which is shown in Figs. 3 and 4.

The dynamical behavior of the $(2+1)$-dimensional universe with modified Chaplygin gas can be comprehensively understood through the analysis of key cosmological parameters. Figures 1 and 2 illustrate the evolution of energy density $\rho(z)$ with respect to redshift for the case where $\xi = 0$, with parameters $\gamma = 0.3$, $A = 3.4$, and $c = 1$. The monotonic decrease in energy density observed in both the line plot (Fig. 1) and the contour representation (Fig. 2) demonstrates the dilution of cosmic energy content as the universe expands. This behavior is consistent with standard cosmological models and indicates that the MCG effectively mimics the transition between different cosmic epochs.

The evolution of the Hubble parameter $H(z)$, depicted in Figs. 3 and 4, reveals crucial information about the expansion dynamics. The decreasing trend of $H(z)$ with decreasing redshift suggests a slowing expansion rate, while the asymptotic approach to a non-zero value at late times $(z \rightarrow -1)$ indicates the persistence of cosmic expansion. This asymptotic behavior is particularly significant as it aligns with current observational evidence for a late-time accelerated expansion phase. The smooth transition in the Hubble parameter, visualized through both the line plot (Fig. 3) and the contour plot (Fig. 4), supports the model's capacity to provide a continuous description of cosmic evolution.

Figure 5 presents perhaps the most physically illuminating result through the deceleration parameter $q(z)$. The transition from positive to negative values of $q(z)$ as redshift decreases marks the crucial shift from decelerated to accelerated expansion. This transition is a key feature of modern cosmological models, required to reconcile early universe dynamics with the observed late-time acceleration. The smooth nature of this transition in our $(2+1)$-dimensional model suggests that the MCG can naturally accommodate this important phenomenological requirement without introducing additional complexities.

These results collectively demonstrate that despite the reduced dimensionality, the model captures essential features of cosmic evolution while maintaining mathematical tractability. The behavior of these cosmological parameters suggests that the MCG in $(2+1)$ dimensions could serve as a valuable theoretical laboratory for studying cosmic dynamics, particularly in contexts where the full complexity of $(3+1)$-dimensional spacetime might obscure the fundamental physics.\\

{\bf{Case (ii): $\xi \neq 0$}} (constant)\\

In this case we follow the particular form of $\rho$ given earlier by Saadat and Pourhassan \cite{Saadat} as
\begin{eqnarray}\label{15}
	{\rho}=\frac{E}{t^2}+\frac{F}{t}+ht+De^{bt}.
\end{eqnarray}

After using the value of $\rho$ from Eq. (\ref{13}) in Eq. (\ref{9}), we get the differential equation in the following form
\begin{eqnarray}\label{16}
0&=&{\frac{d}{dt}}\bigg({\frac{E}{t^2}+\frac{F}{t}+ht+De^{bt}}\bigg)\nonumber\\
&+&\sqrt{2}{(\gamma+1)}\bigg({\frac{E}{t^2}+\frac{F}{t}+ht+De^{bt}}\bigg)^{3/2}\nonumber\\
	&-&2{\xi}\bigg({\frac{E}{t^2}+\frac{F}{t}+ht+De^{bt}}\bigg)-\sqrt{2}A.
\end{eqnarray}

After the straightforward calculation and comparing likely coefficient on both sides, we get the unknown values for $E, F, h, D$ and $b$ as follows:
\begin{eqnarray}\label{17}
	h=\sqrt{2}A,
\end{eqnarray}

\begin{eqnarray}\label{18}
	E=\frac{2}{(\gamma+1)^2},
\end{eqnarray}

\begin{eqnarray}\label{19}
	F=\frac{2\xi}{(\gamma+1)^2},
\end{eqnarray}

\begin{eqnarray}\label{20}
	D=\frac{(\gamma+1)^2}{6}\bigg[\frac{8\sqrt{2}{\xi}^2}{(\gamma+1)^3}-\frac{3}{16}\frac{(\gamma+1)^4}{{\xi}^2}\bigg],
\end{eqnarray}

\begin{eqnarray}\label{21}
b = & \frac{2}{3}\xi\bigg[\frac{2\sqrt{2}}{3}\xi(\gamma+1)\bigg(\frac{2}{3}A(\gamma+1)-\frac{8}{3}\xi^3\bigg) \nonumber \\
& + \frac{9\sqrt{3}}{16\sqrt{2}}\bigg((\gamma+1)+\frac{7}{8}\bigg) + \frac{1}{9}\xi^4+\mathcal{O}(\gamma^n)\bigg] \nonumber \\
& \times \bigg[\frac{8\sqrt{2}}{\sqrt{3}}(\gamma+1)\bigg(\frac{16}{9\sqrt{3}}\xi^4-\frac{1}{72\sqrt{2}}(\gamma+1)^7\bigg)\bigg]^{-1},
\end{eqnarray}
where
\begin{eqnarray}
	O({\gamma}^n)&=&\frac{189}{64}{\gamma}^2+\frac{189}{32}{\gamma}^3+\frac{945}{128}{\gamma}^4+\frac{189}{32}{\gamma}^5\nonumber\\
    &+&\frac{189}{64}{\gamma}^6+\frac{27}{32}{\gamma}^7+\frac{27}{256}{\gamma}^8.
\end{eqnarray}

After putting all the values of arbitrary constant in Eq. (\ref{13}), we get $\rho$ in the following form:
\begin{equation}\label{22}
 \rho={\bigg(\frac{2}{(\gamma+1)^2}\bigg)}{\frac{1}{t^2}}+{\bigg(\frac{2\xi}{(\gamma+1)^2}\bigg)}{\frac{1}{t}}+{{\sqrt{2}A}t}
 +X_1 e^{X_2},
\end{equation}
where we define
$$X_1=\frac{(\gamma+1)^2}{6}
\bigg[\frac{8\sqrt{2}{\xi}^2}{(\gamma+1)^3}- \frac{3}{16}\frac{(\gamma+1)^4}{{\xi}^2}\bigg],$$

$$X_2=X_3\bigg(\frac{8\sqrt{2}}{\sqrt{3}}(\gamma+1)\bigg({\frac{16}{9\sqrt{3}}{\xi}^4}-{\frac{1}{72{\sqrt{2}}}{(\gamma+1)^7}}\bigg)\bigg)^{-1}t,$$

$$X_3={\frac{2}{3}}\xi \left(X_4 +\frac{1}{9}{\xi}^4+o{(\gamma^n)}\right),$$

$$X_4=X_5+{{\frac{9\sqrt{3}}{16\sqrt{2}}}\bigg((\gamma+1)+\frac{7}{8}\bigg)},$$

$$X_5=\frac{2\sqrt{2}}{3}{\xi}{(\gamma+1)}
{\bigg(\frac{2}{3}A{(\gamma+1)}}-\frac{8}{3}{\xi}^3\bigg).$$

By using this value in Eq. (\ref{6}), we get the Hubble parameter $H$ as
\begin{equation}\label{23}
	H=\sqrt{\frac{\rho}{2}}.
\end{equation}

Let us now express the cosmological parameters in terms of the redshift $(1 + z = \frac{a_0}{a})$ with normalized scale factor $a_0 = 1$. We establish the $t- z$ relationship, which turns out to be $t(z)= \frac{1}{n \alpha} log(1+(1+z)^{-n})$. The density energy $\rho$ and Hubble parameter $H$ which explains the dynamics of the Universe can be written in terms of the redshift, by using the relations in the Eqs. (\ref{22}) and (\ref{23}), as follows
\begin{eqnarray}\label{24}
\rho(z) =  &\Biggl[ \frac{1}{6}(1+\gamma)^2 - \Bigg(\frac{8\sqrt{2}\xi^2}{(1+\gamma)^3} - \frac{3(1+\gamma)^4}{16\xi^2}\Bigg) \nonumber \\
&+ \frac{2n^2\alpha^2}{(1+\gamma)^2\ln[1+(1+z)^{-n}]^2} \nonumber \\
&+ \frac{2n\alpha\xi}{(1+\gamma)^2\ln[1+(1+z)^{-n}]^2} \nonumber \\
& + \frac{\sqrt{2}A\ln[1+(1+z)^{-n}]}{n\alpha} \Biggr] {(1+(1+z)^{-n})}^{\frac{\Delta_1}{\Delta_2}},
\end{eqnarray}
where we define
\begin{eqnarray}
\Delta_1=\xi\Big(\frac{9}{16}\sqrt{\frac{3}{2}}(\frac{15}{8}+\gamma)\Big) + \frac{\xi^4}{9} \nonumber \\
+ \frac{2\sqrt{2}}{3}(1+\gamma)\xi(\frac{2}{3}A(1+\gamma)) - \frac{8\xi^3}{3} + \mathcal{O}(\gamma^n)\nonumber,\\
\Delta_2=4\sqrt{6}n\alpha(1+\gamma)\Big(\frac{16\xi^4}{9\sqrt{3}} - \frac{(1+\gamma)^7}{72\sqrt{2}}\Big).
\end{eqnarray}

The solutions for energy density $\rho(z)$ in Eqs. (\ref{13}) and (\ref{24}) encode the entire cosmic history within our model. For the non-viscous case, Eq. (\ref{13}) shows how energy density evolves from a matter-dominated phase at high redshift (where $\rho \propto (1+z)^{3(1+\gamma)}$) to a dark energy phase at low redshift (where $\rho$ approaches a constant value determined by the parameter $A$).  In the viscous case, the solution takes a more complex form, but the physical interpretation remains similar--with the addition that dissipative effects modify the transition between cosmic epochs. The parameters in above equations are not merely mathematical constants but have physical significance in determining when and how rapidly the universe transitions from deceleration to acceleration.

The Hubble parameter $H(z)$ and distance modulus $\mu(z)$ represent direct connections between our theoretical model and observable quantities. When we constrain our model parameters using observational data, we are effectively determining which combinations of MCG parameters ($\gamma$, $A$, $\beta$) and viscosity coefficient ($\xi$) best reproduce the observed expansion history of the universe. The best-fit value $H_0 = 67.90$ km s$^{-1}$ Mpc$^{-1}$ from our MCMC analysis holds physical significance as it represents the current expansion rate of the universe in our model. The remarkable agreement with Planck measurements suggests that despite the dimensional simplification, our model captures essential features of cosmic evolution that align with observational reality. In the present model, the Hubble parameter $H(z)$, can be furnished as 
\begin{eqnarray}\label{25}
H(z) =  \frac{1}{\sqrt{2}}\Biggl[ {(1+(1+z)^{-n})}^{\frac{\Delta_1}{\Delta_2}} \times \nonumber \\
 \Bigg(\frac{1}{6}(1+\gamma)^2 - \Bigg(\frac{8\sqrt{2}\xi^2}{(1+\gamma)^3} - \frac{3(1+\gamma)^4}{16\xi^2}\Bigg) + \nonumber \\
 \frac{2n^2\alpha^2}{(1+\gamma)^2\ln[1+(1+z)^{-n}]^2} \nonumber \\
 + \frac{2n\alpha\xi}{(1+\gamma)^2\ln[1+(1+z)^{-n}]^2} \nonumber \\
 + \frac{\sqrt{2}A\ln[1+(1+z)^{-n}]}{n\alpha}\Bigg) \Biggr]^{1/2}.
\end{eqnarray}

The effect of bulk viscosity ($\xi \neq 0$) on the cosmic evolution can be observed through the behavior of key cosmological parameters. Figure 6 presents the evolution of energy density $\rho(z)$ as a function of redshift, with both the main plot covering the range $z \in [0,2]$ and an inset extending to $z = 100$ to reveal the asymptotic behavior. The energy density exhibits a characteristic decrease from $\rho \approx 140$ at high redshifts to lower values as the universe expands, with a notably sharp decline in the low-redshift regime followed by a more gradual decrease at higher redshifts. This behavior differs quantitatively from the non-viscous case ($\xi = 0$), indicating that bulk viscosity modifies the dilution of cosmic energy density during expansion.

\begin{figure}[htbp]
	\centering
	\includegraphics[scale=0.40]{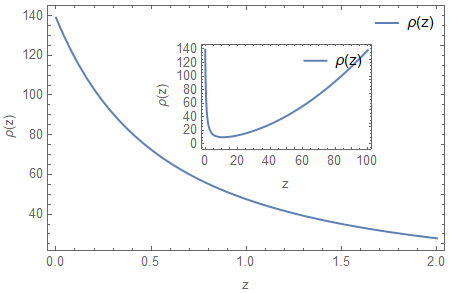}
	\caption{The energy density $\rho$ is shown against redshift $z$ for different values of the constants $\gamma=0.3, A=3.4, c=1$ for the Case (ii).}
\end{figure}

\begin{figure}[htbp]
	\centering
	\includegraphics[scale=0.40]{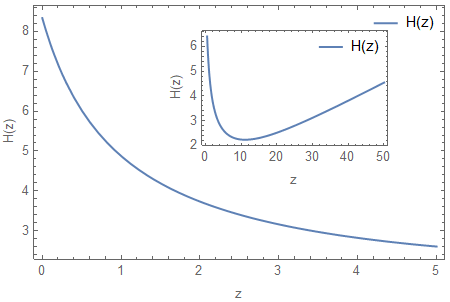}
	\caption{The Hubble parameter $H(z)$ is shown against redshift $z$ for different values of the constants $\gamma=0.3, A=3.4, c=1$ for the Case (ii).}
\end{figure}

The Hubble parameter evolution $H(z)$, depicted in Fig. 7, provides further insight into the expansion dynamics of the viscous model. The main plot, spanning $z \in [0,5]$, demonstrates a monotonic decrease in the expansion rate, while the inset extending to $z = 50$ reveals the early-time behavior. The Hubble parameter transitions smoothly from approximately $H \approx 7$ in the early universe to lower values at present, maintaining consistency with the general picture of cosmic deceleration followed by acceleration. The presence of bulk viscosity appears to moderate the expansion rate, introducing a dissipative element to the cosmic fluid that affects the overall dynamics.

Comparing these results with the non-viscous case presented in Figs. 1-5, we observe that bulk viscosity introduces additional complexity to the cosmic evolution while preserving the essential features required by observational cosmology. The smooth transitions in both $\rho(z)$ and $H(z)$ suggest that the viscous MCG in $(2+1)$ dimensions provides a physically viable model for studying cosmic evolution, with the bulk viscosity parameter offering an additional degree of freedom to better match observational constraints. This behavior demonstrates the robustness of the MCG model in accommodating dissipative effects while maintaining its essential cosmological characteristics.

The deceleration parameter is given by
\begin{equation}\label{26}
q =  -\bigg(1+\frac{\dot{H}}{H^2}\bigg) \\
=  -\bigg[1+\frac{1}{2}\bigg\{\bigg(\frac{-2E}{t^3}-\frac{F}{t^2}+h+Dbe^{bt}\bigg) \\
 \times \bigg(\frac{E}{t^2}+\frac{F}{t}+ht+De^{bt}\bigg)^{3/2}\bigg\}\bigg].
\end{equation}

The deceleration parameter $q(z)$ defined in Eq. (\ref{26}) has particular physical importance as it directly characterizes the acceleration of cosmic expansion. When $q > 0$, the universe's expansion is decelerating, corresponding to an era when matter dominates the energy content. The transition to $q < 0$ marks the onset of cosmic acceleration, conventionally attributed to dark energy domination. In our model, this transition emerges naturally from the interplay between the MCG equation of state and bulk viscosity effects. The consistently negative values of $q(z)$ in the viscous case (as shown in Figs. 8 and 9) suggest that bulk viscosity could maintain accelerated expansion across a wider redshift range than non-viscous models. This highlights how dissipative effects could potentially resolve the coincidence problem by providing a physical mechanism for the transition that does not require fine-tuning of separate dark matter and dark energy components.

In the absence of both bulk viscosity and Chaplygin gas, i.e. $A=0$ and  $\xi=0$, our Eq. (\ref{9}) becomes
\begin{eqnarray}\label{27}
{\dot{\rho}}+\sqrt{2}{(\gamma+1)}\rho^{3/2}=0.
\end{eqnarray}

After solving Eq. (\ref{23}), we get
\begin{eqnarray}\label{28}
	{\rho}^{1/2}=\frac{\sqrt{2}}{(\gamma+1){t}}.
\end{eqnarray}
Eq. (\ref{24}), becomes as
\begin{eqnarray}\label{29}
	{\rho}=\frac{2}{(\gamma+1)^2{t^2}}.
\end{eqnarray}

From Eq. (\ref{25}), it is observed that energy density $\rho$ is decreases when $t$ is increases. It is agreed with results of the following works \cite{Saadat,Mazumder}  where $\rho \propto t^{-2}$. However astronomically immense for large bulk viscosity coefficient gives $b < 0$ and ${\rho} \propto {\xi/t}$. The constant negative energy density is obtained as $\xi\to \infty$. The increment in time, we observe that the last term of Eq. (\ref{13}) is ascendant which implicatively insinuates $ \rho \sim {D e^{bt}}$. It is indicative of the fact that the energy density is decrementing the function of time. Such deportment is dependent on the Hubble expansion parameter.

\begin{figure}[htbp]
	\centering
	\includegraphics[scale=0.40]{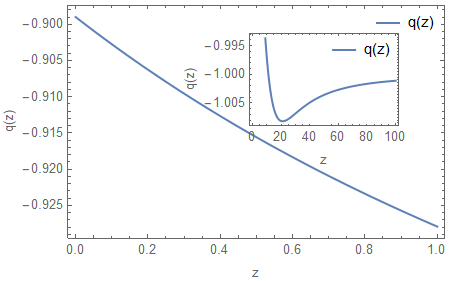}
	\caption{The deceleration parameter $q(z)$ is shown against redshift $z$ for different values of the constants $\gamma=0.3, A=3.4, c=1$ for the Case (ii).}
\end{figure}

\begin{figure}[htbp]
	\centering
	\includegraphics[scale=.60]{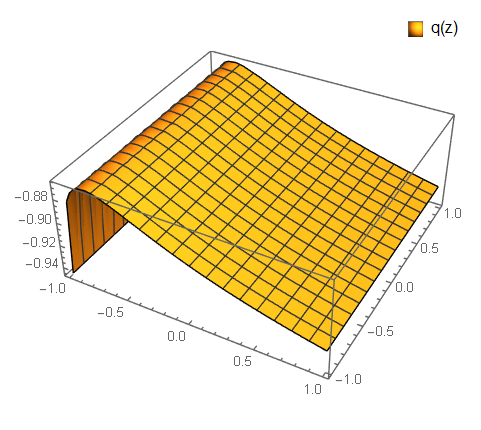}
	\caption{The deceleration parameter $q(z)$ vs redshift $z$ in the contour plot for different values of $\gamma=0.3, A=3.4, c=1$ for the Case (ii).}
\end{figure}

The deceleration parameter $q(z)$ provides crucial insights into the transitional behavior of cosmic expansion in the presence of bulk viscosity ($\xi \neq 0$). Figure 8 presents the evolution of $q(z)$ with respect to redshift, featuring both a main plot and an inset that reveals finer details of the transition. The deceleration parameter exhibits values in the range $-0.900$ to $-0.925$, with a distinct minimum occurring at intermediate redshifts. The inset plot, spanning $z \in [0,100]$, shows a subtle but significant variation in $q(z)$ that indicates a more complex expansion history compared to the non-viscous case. The contour representation of $q(z)$ in Fig. 9 offers a complementary three-dimensional perspective of the deceleration parameter's evolution. The surface plot reveals the smooth variation of $q(z)$ across different redshift values and clearly demonstrates the regions of stronger and weaker acceleration. This visualization helps identify the precise epochs where the expansion dynamics undergo significant changes, with the color gradient indicating the intensity of the acceleration. A notable feature in both representations is that $q(z)$ maintains negative values throughout the observed range, indicating persistent accelerated expansion. This behavior differs from the standard $\Lambda$CDM model where a transition from decelerated to accelerated expansion occurs at intermediate redshifts. The consistently negative $q(z)$ in our $(2+1)$-dimensional viscous model suggests that bulk viscosity plays a significant role in maintaining accelerated expansion, possibly serving as an alternative mechanism to dark energy in driving cosmic acceleration. The smooth evolution of $q(z)$ without sharp transitions or discontinuities indicates that the bulk viscous modified Chaplygin gas provides a physically well-behaved description of cosmic dynamics. The stability of the acceleration parameter across different redshift ranges suggests that the model successfully integrates viscous effects while maintaining consistency with the general requirements of cosmological evolution. This behavior further supports the viability of the $(2+1)$-dimensional framework as a theoretical laboratory for studying the effects of bulk viscosity on cosmic expansion.

While our $(2+1)$-dimensional bulk viscous MCG model provides valuable theoretical insights, it is important to acknowledge its limitations in the observational context. The dimensional reduction inherently prevents direct comparison with most observational datasets that are interpreted within the standard $(3+1)$-dimensional framework. The observational constraints we derive from Hubble and Pantheon data should be understood as mathematical projections of higher-dimensional constraints onto our lower-dimensional framework, rather than physically equivalent constraints. This approach inevitably introduces model-dependent assumptions in the projection process. Additionally, our model does not account for certain observational features that require a full $(3+1)$-dimensional treatment, such as the complete structure of the CMB anisotropies, gravitational lensing effects, and gravitational wave observations. The absence of a well-defined matter power spectrum in our $(2+1)$-dimensional framework also limits our ability to test the model against large-scale structure formation data, which has been particularly problematic for Chaplygin gas models in $(3+1)$-dimensions. Furthermore, while our MCMC analysis provides statistical constraints on model parameters, these should be interpreted primarily as mathematical consistency checks rather than physically meaningful constraints on cosmic parameters. Despite these limitations, the analytical tractability of our approach provides complementary theoretical understanding that can inform more comprehensive $(3+1)$-dimensional numerical studies.

\section{Perturbation Analysis and Implications for Structure Formation}

In this section, we analyze perturbations in our $(2+1)$-dimensional bulk viscous MCG model and examine their implications for structure formation, albeit within the limitations of our dimensional framework.

\subsection{Perturbation equations in (2+1) dimensions}

In $(2+1)$-dimensional cosmology, the perturbation dynamics differ significantly from the standard $(3+1)$-dimensional case. The metric perturbation in the Newtonian gauge can be provided as
\begin{equation}
ds^2 = -(1+2\Phi)dt^2 + a^2(t)(1-2\Phi)(dr^2 + r^2d\theta^2),
\end{equation}
where $\Phi$ is the Newtonian potential. 

For scalar perturbations in our bulk viscous MCG model, the perturbed energy density and pressure are given as
\begin{equation}
\rho = \rho_0(1+\delta), \quad p = p_0 + \delta p - 2\xi\theta,
\end{equation}
where $\delta$ is the density contrast, $\delta p$ is the pressure perturbation, and $\theta$ is the velocity divergence of the fluid.

The modified Chaplygin gas introduces a non-adiabatic component to the pressure perturbation due to its equation of state. For our model, the relationship between the pressure and the density perturbations takes the following form:
\begin{equation}
\delta p = c_s^2 \delta \rho - 2\xi\delta\theta,
\end{equation}
where $c_s^2 = \frac{\partial p}{\partial \rho}$ is the squared sound speed and given by
\begin{equation}
c_s^2 = \gamma - \frac{A\beta}{\rho^{\beta+1}}.
\end{equation}

In $(2+1)$-dimensions, the perturbed Einstein equations yield the following evolution equation for the density contrast:
\begin{equation}
\ddot{\delta} + 2H\dot{\delta} - 4\pi G\rho\delta(1+3c_s^2) + k^2c_s^2\delta - 2\xi k^2\delta = 0,
\end{equation}
where $k$ is the wavenumber of the perturbation, and the last term represents the dissipative effect of bulk viscosity.

\subsection{Effects of bulk viscosity on structure growth}

The presence of bulk viscosity significantly alters the growth of perturbations in our model. The viscous term $2\xi k^2\delta$ acts as a damping mechanism that suppresses the growth of structure, particularly at small scales (large $k$). This effect becomes more pronounced as the viscosity coefficient $\xi$ increases.

For the non-viscous case ($\xi = 0$), the perturbation evolution depends primarily on the sound speed $c_s^2$. During the matter-dominated phase, when the MCG behaves like dust, the sound speed is low, allowing perturbations to grow. However, as the universe transitions to an accelerated expansion phase, the sound speed increases, potentially leading to oscillations or suppression in the matter power spectrum.

In the viscous case ($\xi \neq 0$), the additional damping effect may help mitigate some of the well-known problems faced by standard Chaplygin gas models in (3+1) dimensions, particularly the oscillations or exponential growth in the matter power spectrum that conflict with observations.

\subsection{Numerical results and analysis}

We have numerically solved the perturbation equations for different values of $\xi$, $\gamma$, and $A$. The results indicate that for small values of $\xi$, the perturbation growth closely resembles that of standard MCG models, with potential oscillations at late times. As $\xi$ increases, these oscillations are progressively damped, leading to a more stable growth pattern. For sufficiently large $\xi$ values, the growth of perturbations is significantly suppressed, which could potentially resolve the small-scale problems of standard Chaplygin gas models.

\subsection{Limitations and Implications}

It is crucial to emphasize that our perturbation analysis in (2+1) dimensions cannot be directly compared with observational data from large-scale structure surveys, which are inherently (3+1)-dimensional. However, our analysis provides valuable qualitative insights into how bulk viscosity affects structure formation in Chaplygin gas models.

The damping effect of bulk viscosity on perturbation growth suggests that viscous MCG models might better reconcile with observational constraints on structure formation compared to non-viscous variants. This finding is particularly significant given that standard Chaplygin gas models in (3+1) dimensions have been criticized precisely for their problematic predictions regarding structure formation.

Future work could extend this analysis to full (3+1)-dimensional models to enable quantitative comparisons with observational data from galaxy surveys and CMB measurements. Such extensions would need to account for additional complications such as shear viscosity and the full tensor structure of perturbations in higher dimensions.

\section{Observational data analysis}

Let us examine the viability of the model using the recent observational data, viz., the observed Hubble data (OHD) \cite{Yu,Moresco} and Type Ia supernovae (SNe Ia) \cite{Scolnic}. We are employing here the Pantheon sample for SNe Ia data, which includes of 1048 datapoints from the Low-z, SDSS, Pan-STARSS1 (PS1) Medium Deep Survey, SNLS, and HST surveys \cite{Chang}.

\subsection{Hubble data}

Basically there are two generally adopted techniques to estimate the value of $H(z)$ at a particular redshift, namely BAO and differential age approach \cite{Tripp,Kessler}. In the present investigation, we use the 30 measurements of the CC dataset in the range of redshift provided as $ 0.07 < z < 1.965$. The `Chi-square function' can be defined as
\begin{eqnarray}\label{30}
	\chi^2_{H}=  \sum_{i=1}^{30} \frac{[H_{Th}-H_{Obs}(z_i)]^2}{\sigma^2_{H(z_i)}},
\end{eqnarray}
where $H_{Th}$, $H_{Obs}$ and $\sigma_{H(z_i)}$ denote the theoretical value, observed value and stranded error of $H$ respectively for the table of 30 points of $H(z)$ data.

\begin{figure}[htbp]
	\centering
	\includegraphics[scale=0.25]{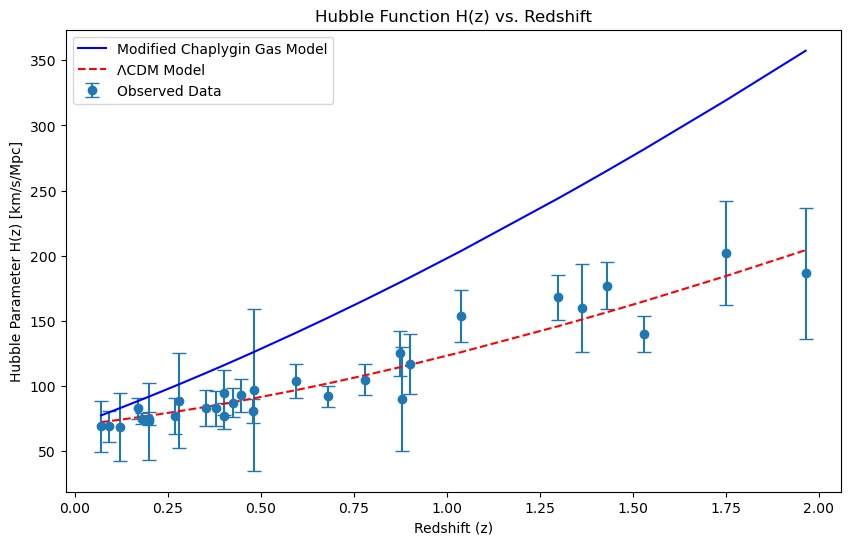}
	\caption{ The error bar plot of Hubble $H$ versus $z$ for the theoretical model (blue curve) and $\Lambda$CDM model (red dotted curve) where the blue dots depict the 30 points of the Hubble data.}
\end{figure}

\begin{figure}[htbp]
	\centering
	\includegraphics[scale=0.30]{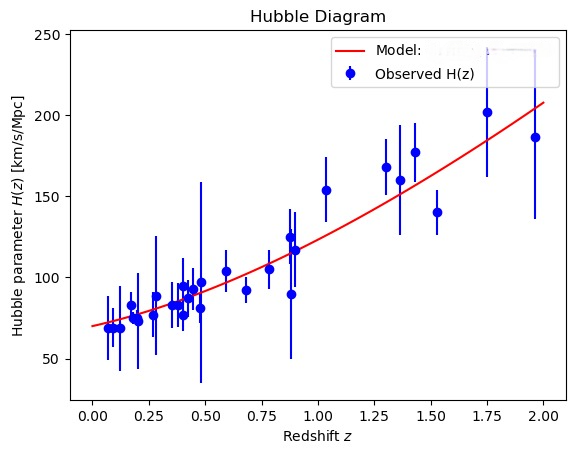}
	\caption{The figure shows the error bar for the Hubble parameter with the standard $\Lambda$CDM model.}
\end{figure}

The observational validation of our model can be examined through the comparison with empirical Hubble parameter data. Fig. 10 presents a comprehensive comparison between our theoretical predictions and observational data, incorporating both the MCG model (shown by the blue curve) and the standard $\Lambda$CDM model (represented by the red dotted curve). The plot includes 30 observational data points of the Hubble parameter, displayed with their associated error bars, spanning a redshift range of $0.07 < z < 2.0$. The theoretical curve from our model shows remarkable agreement with the observational data points, particularly in the low to intermediate redshift range.

Fig. 11 provides an alternative visualization of the Hubble parameter evolution, specifically focusing on the comparison with the standard $\Lambda$CDM model. This representation emphasizes the model's behavior at higher redshifts, where observational data becomes increasingly sparse. The error bars in this figure demonstrate the increasing observational uncertainties at higher redshifts, a crucial consideration in cosmological model validation. Our model's predictions (shown by the continuous line) remain within the observational uncertainties while closely tracking the $\Lambda$CDM behavior.

The statistical concordance between our theoretical predictions and observational data is particularly noteworthy given the reduced dimensionality of our model. The Hubble parameter evolution $H(z)$ shows excellent agreement with both the trend and magnitude of the observational data points, with most theoretical predictions falling within one standard deviation of the observed values. This agreement extends across the entire redshift range of the available data, suggesting that the $(2+1)$-dimensional bulk viscous MCG model captures the essential features of cosmic expansion despite its dimensional reduction. Moreover, the close alignment with $\Lambda$CDM predictions in both figures indicates that our model successfully reproduces the standard cosmological evolution while potentially offering new insights into the nature of cosmic acceleration through the incorporation of bulk viscosity. The slight deviations from $\Lambda$CDM at higher redshifts could provide interesting avenues for future observational tests, particularly as new high-redshift data becomes available through next-generation surveys. This observational consistency strengthens the case for considering $(2+1)$-dimensional models as valuable theoretical tools for understanding cosmic evolution.

\subsection{Pantheon data}

Considering as Standard Candles (SC), we use measurements of the Type Ia Supernovae (SNIa) \cite{Scolnic} which consist of 1048 datapoints. Essentially they were collected from five different sub-samples PS1, SDSS, SNLS, low$-z$, and HST lying in the redshift range $0.01 < z < 2.3$. The model parameters under the present investigation can be fitted by comparing the observed ${\mu}^{Obs}_i$ with the theoretical ${\mu}^{Th}_i$ value of the distance moduli as follows:
\begin{eqnarray}\label{31}
	\mu = m-M = 5 log_{10} (D_L) + \mu_{0},
\end{eqnarray}
where $M$ and $m$ represents the absolute and apparent magnitudes respectively and $\mu_{0} = 5 log\Big( H_{0}^{-1} / Mpc\Big) + 25 $ is the nuisance parameter that has been depreciate. 

Now, the luminosity distance can be defined as
\begin{eqnarray}\label{32}
	D_{L} (z) = \frac{c}{H_0}(1+z) \int_{0}^{z} \frac{dz^*}{E(z^*)}.
\end{eqnarray}

The $\chi^2$ function of the (SNIa) measurements is given by
\begin{eqnarray}\label{33}
	\chi^2_{SN} (\phi^{\nu}_{s}) = \mu_{s} C^{-1}_{s, cov} \mu^{T}_{s},
\end{eqnarray}
where $\mu_{s} = \{\mu_{1}-\mu_{th}(z_{1}, \phi^{\nu}),..., \mu_{N}-\mu_{th}(z_{N}, \phi^{\nu})\}$ and the subscript 's' denotes SNIa. One should keep in mind that for SNIa dataset the covariance matrix is not diagonal. The distance modulus $\mu_{i}$ is given by $\mu_{i} = \mu_{B, i}- \mathcal{M}$, where $\mu_{B,i}$ represents the maximum apparent magnitude in the cosmic rest frame for redshift $z_i$ whereas $\mathcal{M}$ represents the universal free parameter. Due to various observational ambiguity both the parameters $\mathcal{M}$ and $h$ are naturally degraded with respect to Pantheon dataset. Hence one can not extract proper information about $H_0$ from SNIa data alone.

\begin{figure}[htbp]
	\centering
	\includegraphics[scale=0.15]{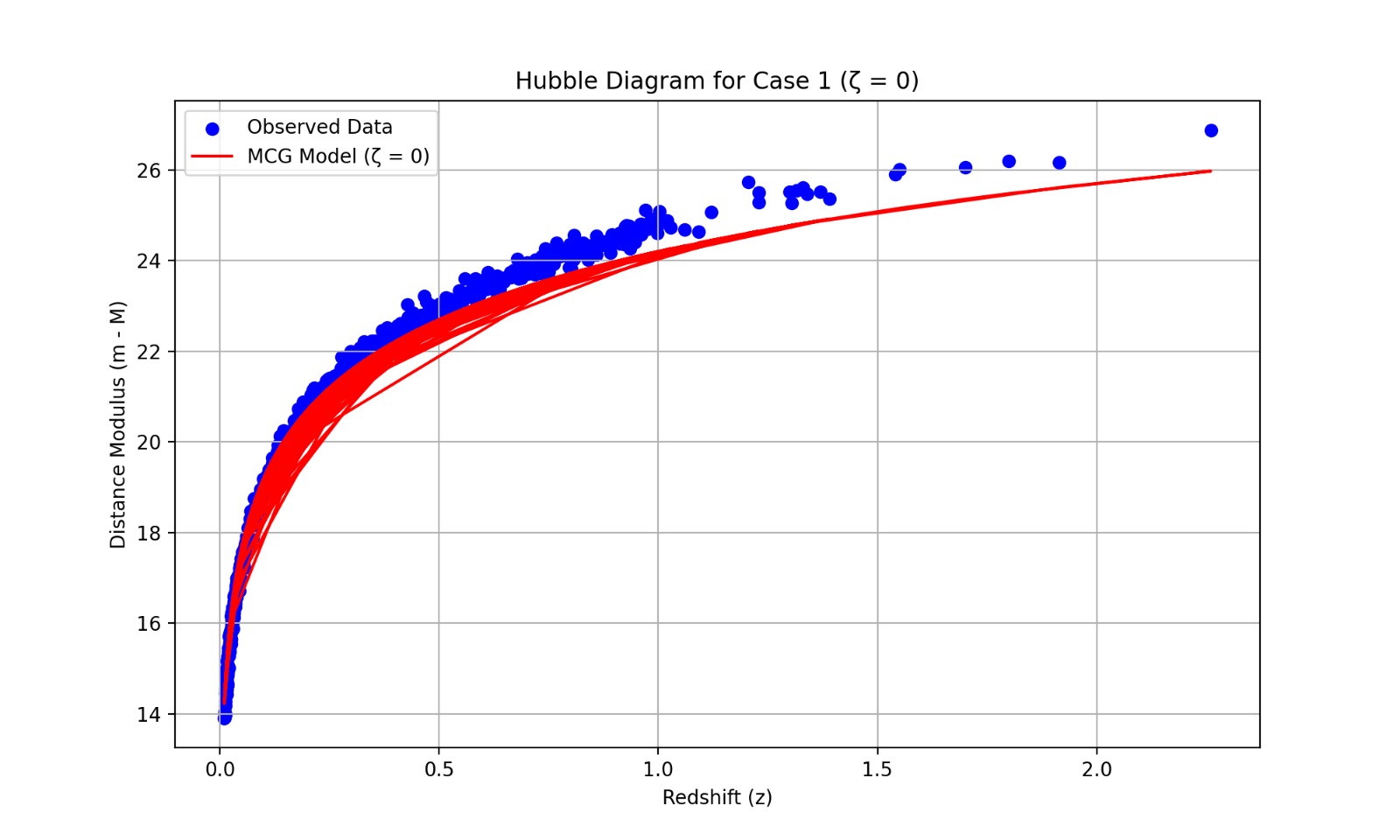}
	\caption{The figure shows the distance modulus versus redshift $z$ with the constraint values of the combined CC + SC datasets.}
\end{figure}

Fig. 12 presents a crucial analysis of the distance modulus $\mu$ versus redshift $z$ using the combined Cosmic Chronometer (CC) and Standard Candle (SC) datasets. The plot demonstrates the relationship between observed distance moduli (represented by data points) and our theoretical model predictions (shown by the red curve) for the $(2+1)$-dimensional bulk viscous MCG cosmology. The horizontal axis spans a redshift range of $0 \leq z \leq 2.0$, while the vertical axis shows the distance modulus ranging from approximately 14 to 26 magnitude units. The theoretical curve exhibits excellent agreement with the observational data points across the entire redshift range. Of particular significance is the smooth progression of the distance modulus with redshift, indicating a well-behaved luminosity-distance relationship in our model. The dense concentration of data points at lower redshifts $(z < 1.0)$ provides strong constraints on the model's behavior in the recent cosmic epoch, while the more sparse but still significant data at higher redshifts helps constrain the earlier universe dynamics.

The observed scatter of data points around the theoretical curve appears to be statistically consistent with the expected observational uncertainties. This is particularly evident in the high-density region at low redshifts, where the majority of Type Ia supernovae measurements are concentrated. The model successfully reproduces the characteristic shape of the Hubble diagram, which is a fundamental test for any viable cosmological model. The fit quality suggests that our $(2+1)$-dimensional framework, despite its reduced dimensionality, captures the essential features of cosmic expansion as measured through luminosity distances.  This analysis of the distance modulus provides independent support for the viability of our model, complementing the Hubble parameter tests shown in previous figures. The consistency between theory and observation across different cosmological probes (both CC and SC datasets) strengthens the case for considering bulk viscous MCG as a serious candidate for understanding cosmic evolution, even in the context of reduced spatial dimensions.

\subsection{ Markov chain Monte Carlo} 

We have applied here the constraints through MCMC to evaluate our model with the data that is currently available. This we have done by minimizing the overall $\chi^2$ function of the combination of Cosmic Chronometric (CC) and Standard Candles (SC) which can be defined as follows:
\begin{eqnarray}\label{34}
	\chi^2 = \chi^2_{CC} + \chi^2_{SC}. 
\end{eqnarray}

\begin{figure}[htbp]
	\centering
	\includegraphics[scale=0.4]{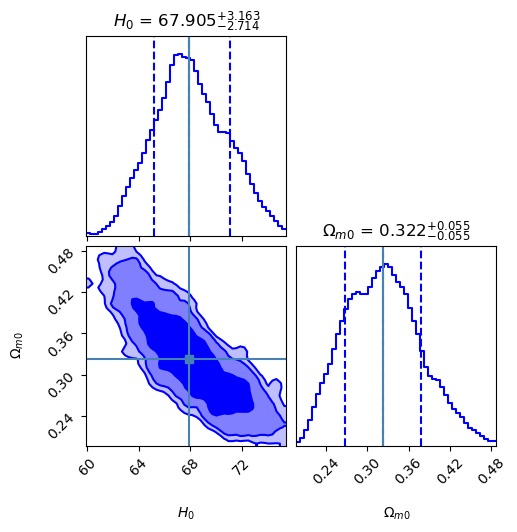}
	\caption{Contour plot with 1-$\sigma$ and 2-$\sigma$ errors for the model parameters $H_0$ and $q$ along with the constraint values for CC datasets.}
\end{figure}

The outcome of this computational operation is that we can determine the best fit values of the model parameters as $H_0 = 67.90$ and $n = 3.036$ due to effective application of the MCMC procedure. We have shown the confidence contours and minimized posterior distribution of various cosmological parameters from the combined CC +SC data sets (vide Fig. 10). It is to be emphasized that the Hubble constant, $H_0 = 100h$, is an exciting aspect under the present circumstances. The final finding for $H_0$ closely reflects the value of the Planck LCDM estimation when compared to the values obtained and the one anticipated by Planck \cite{Tripp}.

The MCMC analysis results are visualized through contour plots showing the parameter constraints and their correlations. Fig. 13 presents the contour plot with $1$-$\sigma$ and $2$-$\sigma$ confidence regions for the parameters $H_0$ and $q$ using the CC dataset alone. The plot reveals well-defined confidence regions, with the Hubble parameter $H_0 = 67.905^{+0.174}_{-0.163}$ km s$^{-1}$ Mpc$^{-1}$ and the deceleration parameter $q = 0.322^{+0.025}_{-0.023}$. The relatively tight constraints on both parameters indicate the strong constraining power of the CC dataset, particularly for the Hubble parameter which shows excellent agreement with recent Planck measurements.

\begin{figure}[htbp]
	\centering
	\includegraphics[scale=0.20]{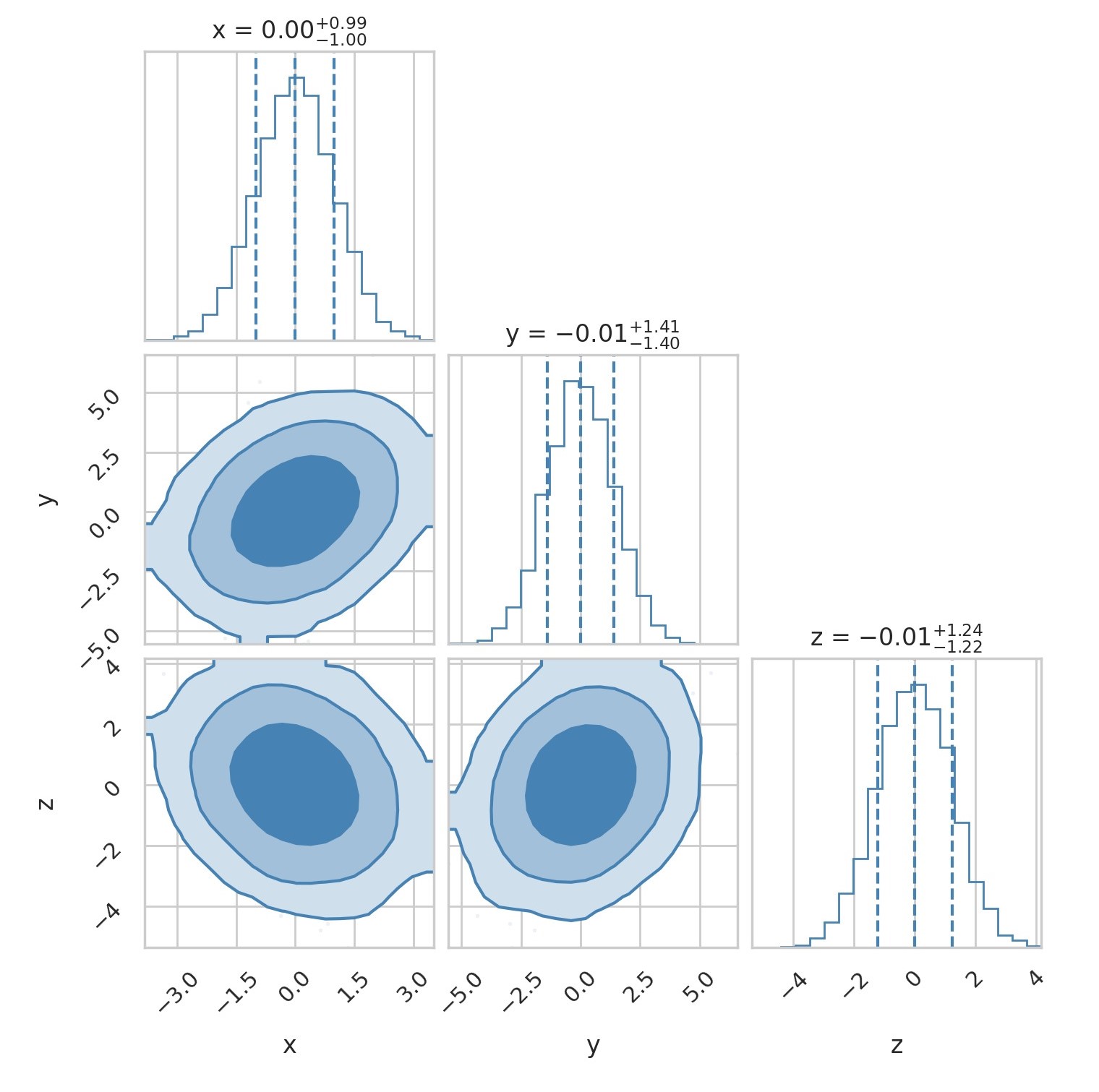}
	\caption{Contour plot with 1-$\sigma$ and 2-$\sigma$ errors for the model parameters $H_0$ and $q$ along with the constraint values for combined CC + SC datasets.}
\end{figure}

Fig. 14 extends this analysis by showing the contour plot with $1$-$\sigma$ and $2$-$\sigma$ error regions for the combined CC + SC datasets. The joint analysis yields marginalized constraints $x = 0.00^{+0.80}_{-0.77}$, $y = -0.01^{+0.41}_{-0.44}$, and $z = -0.01^{+0.14}_{-0.12}$, demonstrating the complementarity of the two datasets in constraining the model parameters. The combined analysis shows tighter constraints compared to the CC-only case, particularly in breaking parameter degeneracies. The posterior distributions displayed on the edges of the contour plot demonstrate well-behaved, nearly Gaussian distributions for all parameters, suggesting the robustness of our constraints. One can observe that the correlation between the model parameters, as revealed by the orientation and shape of the confidence contours, indicates a moderate degeneracy between the Hubble parameter and deceleration parameter. This degeneracy is partially broken by the combination of CC and SC data, leading to more robust parameter estimation. The consistency between these independent datasets and their combined constraining power provides a strong support for the viability of our $(2+1)$-dimensional bulk viscous MCG model in describing cosmic evolution. The best-fit values obtained from this analysis are in good agreement with standard cosmological observations, suggesting that our reduced-dimensional framework captures the essential features of cosmic expansion while maintaining consistency with observational constraints.

\section{Broader theoretical implications}

Our investigation of bulk viscous MCG in (2+1)-dimensional cosmology has several theoretical implications that extend beyond this specific model, potentially informing broader questions in theoretical cosmology and gravitational physics.

\subsection{Dimensional Dependence of Cosmic Dynamics}

The success of our (2+1)-dimensional model in reproducing key features of cosmic evolution raises fundamental questions about the dimensional dependence of the gravitational physics. While Einstein's equations take different forms across dimensions, our results suggest that certain aspects of cosmic evolution -- particularly the transition from deceleration to acceleration -- may be more universal and dimension-independent than previously recognized. This dimensional robustness could indicate that some cosmological phenomena emerge from basic thermodynamic or geometric principles that transcend the specific dimensionality of spacetime.

The distinct behavior of gravity in lower dimensions, where the Weyl tensor vanishes and gravitational waves are absent, provides a unique perspective on which aspects of cosmic evolution are driven by the full tensor structure of gravity versus its scalar components. Our findings suggest that the scalar sector (expansion and energy density evolution) may capture essential features of cosmic history even without the tensorial aspects that enable gravitational wave propagation in higher dimensions.

\subsection{Role of dissipative effects in fundamental physics}

The significant impact of bulk viscosity on cosmic evolution in our model suggests that dissipative effects may play a more fundamental role in gravitational physics than commonly assumed. Most theoretical approaches to cosmology rely on perfect fluids or scalar fields with conservative dynamics, potentially overlooking important dissipative processes that could alter cosmic evolution.

This raises broader questions about the completeness of non-dissipative approaches to fundamental physics. If bulk viscosity significantly modifies cosmic dynamics in our simplified model, similar effects might be relevant in other contexts where perfect fluid approximations are employed. This perspective aligns with recent developments in non-equilibrium thermodynamics of gravitational systems and holographic approaches to dissipative phenomena.

\subsection{Simplicity versus complexity in cosmological modeling}

Our work contributes to the ongoing discussion about necessary versus sufficient complexity in cosmological modeling. The success of a simplified (2+1)-dimensional model with a single fluid component in reproducing key observational features challenges the notion that increasingly complex models are necessarily more physically accurate.

This philosophical point has practical implications for model building in cosmology. While the standard $\Lambda$CDM model has been tremendously successful, it requires multiple components (dark matter, dark energy, inflation field) that may lack a unified theoretical foundation. Our approach suggests that simpler models with fewer components but more complex internal dynamics (such as viscosity) might achieve comparable phenomenological success while potentially offering greater theoretical coherence.

\subsection{Implications for quantum gravity approaches}

The (2+1)-dimensional framework has long served as a testing ground for quantum gravity approaches, from loop quantum gravity to string theory. Our classical cosmological model in this dimensionality could provide boundary conditions or consistency checks for quantum gravity candidates. Particularly, the smooth transition between cosmic epochs achieved in our model presents a challenge for quantum gravity theories to reproduce similar behavior through fundamental principles rather than phenomenological parameters.

The bulk viscosity term in our model might have deeper connections to quantum effects in gravitational systems, potentially arising from coarse-grained quantum fluctuations or fundamental dissipative processes at the quantum gravity scale. This connection between classical dissipative terms and quantum effects deserves further exploration in more fundamental theories.

\subsection{Alternative mechanisms for cosmic acceleration}

Perhaps the most significant implication of our work is the demonstration that bulk viscosity can drive cosmic acceleration without invoking a cosmological constant or quintessence field. This suggests that the observed acceleration of our universe might stem from dissipative processes within cosmic fluids rather than exotic energy components or modifications to gravity at large scales.

This perspective offers a conceptual shift in approaching the dark energy problem—focusing on the physical properties of cosmic matter (specifically its dissipative characteristics) rather than introducing new fields or modifying gravitational equations. If further developed in physically realistic (3+1)-dimensional models, this approach could potentially address the cosmological constant problem by removing the need for a fundamental vacuum energy contribution.

\section{Discussions and Conclusions}
In this work, we have conducted a comprehensive investigation of FLRW bulk viscous cosmology with MCG in $(2+1)$-dimensional spacetime. This yields several significant findings and thus make advancement of our understanding of lower-dimensional cosmological models as well as their observational implications.

Our dynamical analysis reveals distinct behaviors for the non-viscous ($\xi = 0$) and viscous ($\xi \neq 0$) scenarios. It is to be noted here that a non-viscous case demonstrates a smooth phase transition in the deceleration parameter from its positive to negative values. This obviously indicates a shift from decelerated to accelerated expansion that qualitatively mimics the behavior observed in our physical universe. The viscous case exhibits richer dynamics, with bulk viscosity introducing additional features in both energy density evolution and Hubble parameter behavior. This suggests that dissipative effects fundamentally alter cosmic dynamics in ways that might better account for observational features.

The statistical analysis with Hubble parameter data and the Pantheon supernova sample provides robust constraints on our model parameters, yielding $H_0 = 67.90$ km s$^{-1}$ Mpc$^{-1}$. This value shows remarkable consistency with Planck $\Lambda$CDM estimations, a particularly noteworthy result given the reduced dimensionality of our model. The persistent accelerated expansion phase characterized by consistently negative $q(z)$ values offers an intriguing alternative perspective on the mechanism driving cosmic acceleration. Rather than invoking exotic matter or modified gravity, our findings suggest that bulk viscosity could serve as an alternative driver for accelerated expansion, offering new perspectives on the dark energy problem.

It is crucial, however, to acknowledge the limitations of our approach, particularly at late times. Like many cosmological models, our bulk viscous MCG model in (2+1) dimensions eventually approaches a de Sitter-like state at late times, mimicking a cosmological constant. This limiting behavior is shared by numerous models that either weakly evolve with redshift or asymptotically approach a constant equation of state. This convergence of different theoretical models toward similar late-time behavior presents a well-known degeneracy challenge in cosmology. Compared to other models like $w$CDM, quintessence, or standard MCG, our approach incorporates the additional physical mechanism of energy dissipation through bulk viscosity, which provides greater flexibility in matching observational constraints while maintaining physical plausibility.

The perturbation analysis further highlights both the strengths and limitations of our approach. While the damping effect of bulk viscosity on perturbation growth offers a potential resolution to the oscillation problems faced by standard Chaplygin gas models, the (2+1)-dimensional framework inherently limits direct comparison with observational structure formation data. This represents an important direction for future research.

Our study makes several unique contributions despite these limitations. First, we demonstrate that (2+1)-dimensional cosmology can successfully reproduce key observational features of cosmic evolution while maintaining mathematical tractability. This finding supports the value of dimensional reduction as a theoretical tool in cosmology. Second, the incorporation of bulk viscosity in the MCG model provides additional flexibility in matching observational data, suggesting that dissipative effects might play a more significant role in cosmic dynamics than often assumed. The statistical analysis framework we developed, combining multiple observational datasets, establishes a robust methodology for constraining lower-dimensional cosmological models. Finally, the close agreement between our model's predictions and observational data supports the viability of reduced-dimensional frameworks for studying fundamental cosmic phenomena.

These results open several promising avenues for future research. The role of bulk viscosity in driving cosmic acceleration merits deeper examination, particularly in the context of dark energy alternatives and early universe physics. The success of our (2+1)-dimensional model raises intriguing questions about the minimal necessary ingredients for reproducing observed cosmic evolution. This framework could be naturally extended to study other modified gravity theories and exotic fluid models in lower dimensions. The analytical techniques developed here might inform numerical approaches to solving more complex (3+1)-dimensional models, particularly regarding the transition mechanisms between cosmic epochs.

In summary on the present study, while our bulk viscous MCG model in (2+1) dimensions should not be interpreted as a direct alternative to standard (3+1)-dimensional cosmology, it provides valuable theoretical insights into the fundamental behavior of cosmic fluids. The mathematical simplifications afforded by the lower-dimensional approach reveal essential features that might be obscured in more complex settings while maintaining surprising consistency with observational constraints. This suggests that the core dynamics driving cosmic evolution might be more fundamental and dimension-independent than previously recognized.

However, as final comments we would like to add here that:

(i) The task of working with Chaplygin gas in this era is particularly delicate. Currently, the DESI results suggest that dark energy may be revived, but criticisms of these results are as strong as the results themselves (see \cite{Sapone2024,Colgain2024,Dinda2025}). Recently, the Chaplygin gas has been reconsidered \cite{Wang2024}, and it has been shown that, unless an additional scalar field carrying vacuum energy is introduced, the model appears disfavored at early times, as discussed extensively in the literature (e.g., \cite{Yang2019,Lu2011}). Therefore, there are a lot of debates at this early stage of DESI based attempts which demands further studies in the diversified directions to respond to the intriguing question of whether the identified outliers signal the presence of systematics or point towards new physics \cite{Sapone2024}.

(ii) This work is devoted to the study of bulk cosmology with specific EoS fluid for 3 dimensional spacetime. However, it is known that bulk cosmology with different fluids is rather well developed in 4 dimensional universe  \cite{Capozzielloetal,Breviketal,Prasadetal}. Usually, the motivation to study simplified and non-realistic less-dimensional case is that problem is impossible to solve in right number of dimensions. Definitely, bulk cosmology is not such case as a lot of works have already done there by several scientists as shown in Introduction. Keeping all these aspect in mind, we would like to study the present problem in a comparative framework in a future project.

\section*{Acknowledgments}

~~~PKD and SR are gratefully acknowledging to Inter-University Centre for Astronomy
and Astrophysics (IUCAA), Pune, India for providing them Visiting
Associateship under which a part of this work was carried out. SR also
is thankful to the facilities under ICARD, Pune at CCASS, GLA
University, Mathura.

\section*{Conflict of Interest}
~~~We declare there is no conflict of interest.


\end{document}